\def\tilde{\widetilde}
\def\bar{\overline}
\def\hat{\widehat}
\def\*{\star}
\def\[{\left[}
\def\]{\right]}
\def\({\left(}
\def\){\right)}

\def\frac#1#2{{#1 \over #2}}
\def\inv#1{{1 \over #1}}
\def\half{{_1\over^2}}
\def\d{\partial}

\def\vev#1{\langle #1 \rangle}

\def\2pi{\hbox{$2\pi i$}}

\def\dsl{\raise.15ex\hbox{/}\kern-.57em\partial}
\def\Dsl{\,\raise.15ex\hbox{/}\mkern-.13.5mu D}

\def\s{\hspace{0.05cm}}
         \def\Ga{\Gamma}

\def\al{\alpha}
\def\ep{\epsilon}
\def\la{\lambda}        
\def\de{\delta}         \def\De{\Delta}
         
\def\sig{\sigma}        

\def\CM{{\cal M}}              \def\CO{{\cal O}}

\font\numbers=cmss12
\font\upright=cmu10 scaled\magstep1
\def\stroke{\vrule height8pt width0.4pt depth-0.1pt}
\def\topfleck{\vrule height8pt width0.5pt depth-5.9pt}
\def\botfleck{\vrule height2pt width0.5pt depth0.1pt}
\def\Zmath{\vcenter{\hbox{\numbers\rlap{\rlap{Z}\kern
		0.8pt\topfleck}\kern
		2.2pt \rlap Z\kern 6pt\botfleck\kern 1pt}}}
\def\Qmath{\vcenter{\hbox{\upright\rlap{\rlap{Q}\kern
		   3.8pt\stroke}\phantom{Q}}}}
\def\Nmath{\vcenter{\hbox{\upright\rlap{I}\kern 1.7pt N}}}
\def\Cmath{\vcenter{\hbox{\upright\rlap{\rlap{C}\kern
		   3.8pt\stroke}\phantom{C}}}}
\def\Rmath{\vcenter{\hbox{\upright\rlap{I}\kern 1.7pt R}}}
\def\Z{\ifmmode\Zmath\else$\Zmath$\fi}
\def\Q{\ifmmode\Qmath\else$\Qmath$\fi}
\def\N{\ifmmode\Nmath\else$\Nmath$\fi}
\def\C{\ifmmode\Cmath\else$\Cmath$\fi}
\def\R{\ifmmode\Rmath\else$\Rmath$\fi}
\def\cadremath#1{\vbox{\hrule\hbox{\vrule\kern8pt\vbox{\kern8pt
			\hbox{$\displaystyle #1$}\kern8pt}
			\kern8pt\vrule}\hrule}}

\def\presentation{
\voffset -.50in   
\hoffset -.19in
\oddsidemargin 0in \evensidemargin 0in
\marginparwidth .75in \marginparsep 7pt \topmargin 0in
\headheight 12pt \headsep .25in
\footheight 18pt \footskip .35in
\textheight 9.5in \textwidth 6.5in
\columnsep 10pt \columnseprule 0pt }
\def\qq{ \begin{eqnarray} }
\def\qqq{ \end{eqnarray} }
\def\non{ \nonumber }
\documentstyle[12pt]{article}
\presentation
\begin{document}
\vskip 1cm
\centerline{\LARGE Anomalous scaling in the $N$-point}
\bigskip
\centerline{\LARGE functions of passive scalar}
\vskip 1cm
\vskip2cm
\centerline{\large  D. Bernard${}^{a,b}$,\hspace{-0.25cm}
\footnote[1]{Membre du CNRS} \ K.
Gaw\c{e}dzki${}^{a}{}^{1}$ \ and \ A. Kupiainen${}^{c}$}
\ \

\medskip

\centerline{${}^a$ Institut des Hautes Etudes Scientifiques}
\centerline{F-91440, Bures-sur-Yvette, France.}
\medskip
\centerline{${}^{b}$ Service de Physique Th\'eorique de Saclay
\footnote[2]{Laboratoire de la Direction des Sciences de la
Mati\`ere du Commisariat \`a l'Energie Atomique.}}
\centerline{F-91191, Gif-sur-Yvette, France.}
\medskip
\centerline{${}^c$ Mathematics Department, Helsinki University}
\centerline{PO Box 4, 00014 Helsinki, Finland}
\vskip 3cm
\centerline{\large Abstract}
\vskip 1cm
\noindent A recent analysis of the 4-point correlation
function of the passive scalar advected by a time-decorrelated
random flow is extended to the $N$-point case. It is shown
that all stationary-state inertial-range correlations are
dominated by homogeneous zero modes of singular operators
describing their evolution. We compute analytically the zero
modes governing the $N$-point structure functions and
the anomalous dimensions corresponding to them to the linear
order in the scaling exponent of the 2-point function
of the advecting velocity field. The implications
of these calculations for the dissipation correlations
are discussed.

\vfill
\newpage
%
%
\section{Introduction}

There has been much effort lately to understand
the behavior of a scalar quantity passively advected
by a random flow with a Gaussian statistics decorrelated
in time \cite{Krai}. This simple model, of its own
interest, has served as a prototype of a turbulent
system. It is believed that its behavior may teach us
important lessons about the fully developed hydrodynamical
turbulence. One of the interesting aspects of the
passive scalar which has been recently understood
\cite{KG,Weiz,Shrai} is the origin of the breakdown
of Kolmogorov inertial-range scaling in the higher
structure functions of the scalar. It has been
realized that the dominant contribution to the structure
functions comes from the zero modes of the differential
operators describing the stochastic evolution of the
correlation functions of the scalar. In this note,
we extend the results of ref.\s\s\cite{KG} by presenting
the computation of the anomalous dimensions of
the $N$-point structure functions in the first order
of the parameter $\xi$. Exponent $\xi$, which
in \cite{KG} was denoted $\kappa$ and in \cite{Weiz}
$2-\gamma$, is the growth rate of the 2-point structure
function of the velocities of the advecting flow.
The present work was motivated by \cite{Weiz2}
where a similar analysis in the first order in inverse
dimension was sketched.
\vskip 0.3cm

The equation governing the passive scalar in a turbulent
flow is:
\qq
\d_tT + (u\cdot\nabla)T -  \nu \De T = f\s.
\label{eqT}
\qqq
Here $T(x,t)$ describes the scalar, e.g. the temperature,
and $f$ the forcing term whose role is
to compensate the dissipation caused
by the term proportional to the molecular diffusivity $\nu$.
The velocity field $u$  with $\nabla\cdot u=0$ is supposed
to be random. We shall work in $d\geq3$ space dimensions
and shall assume homogeneity, isotropy and parity invariance
of the advecting flow and of the forcing.

\vskip 0.3cm

The statistics of the forcing term is assumed to be Gaussian with
mean zero and 2-point function
\qq
\vev{f(x,t)f(y,t')}= C({\frac{_{x-y}}{^L}})\s\de(t-t')
\label{vevf}
\qqq
The rotation-invariant function $C(x/L)$, which could
be chosen to be a Gaussian, varies on scale $L$.
\vskip 0.3cm

The statistics of the velocity field, independent
of the forcing, is also supposed to be
Gaussian with zero mean and with the 2-point functions
\qq
\vev{u^\alpha(x,t)u^\beta(y,t')}= D^{\alpha\beta}(x-y)\de(t-t')
\quad {\rm with}\quad \d_{\alpha}D^{\alpha\beta}=0.
\label{vevu}
\qqq
To analyze the scaling property of the scalar correlation
functions we shall use the following expression for
$D^{\alpha\beta}$\s:
$D^{\alpha\beta}(x)= D(0)\de^{\alpha\beta}
- d^{\alpha\beta}(x)$ with
\qq
d^{\alpha\beta}(x)=
D\({ (d+\xi -1)\s \delta^{\alpha\beta}-
\xi \frac{x^\alpha x^\beta}{|x|^2}}\)|x|^\xi
\label{defD}
\qqq
where $\xi$ is a parameter, $0<\xi<2$.
Clearly, this distribution for $u$ is far from realistic.
It mimics however the growth of the correlations of
velocity differences with separation distance, typical
for turbulent flows. The fact that the 2-point functions
(\ref{vevf},\ref{vevu}) are white noise in time
is crucial for the solvability of the model.
The parameter $\xi$ fixes the naive dimensions
under the rescalings $x\to \mu x$, $L\to\mu L$.  The naive
dimension of $u$ is $\xi/2$ and of $T$ is $(2-\xi)/2$.
Scale $L$ serves as an infrared cutoff and
the ``Kolmogorov scale" $\eta=\({\frac{\nu}{D}}\)^{1/\xi}$
as an ultraviolet cutoff.
\vskip 0.3cm

We shall be interested in the correlation functions of the
scalar in the inertial range $\eta\ll x \ll L$.
The main result of this note is that in this range
the stationary-state, equal-time, even structure functions
scale with the anomalous exponents $\rho_N$ as
\qq
\vev{\({T(x,t)-T(0,t)}\)^N}\ \cong\ a_N
\({\frac{L}{|x|}}\)^{\rho_N}~|x|^{(2-\xi)N/2}
\label{behav}
\qqq
with
\qq
\rho_N=\xi\frac{N(N-2)}{2(d+2)} + \CO(\xi^2)\s.
\label{anormal}
\qqq
The exponents are universal depending only on
$\xi$ but the amplitudes $a_N$ are not: they
depend on the shape of the covariance $C$.
The error term is bounded by $\CO((L/|x|)^{-2+\CO(\xi)}
|x|^{(2-\xi)N/2})$ so it is strongly
suppressed for large $L/|x|$. As it should be,
the $\rho_N$'s satisfy the H\"{o}lder inequality
$\rho_N \geq \frac{N-2}{2}\rho_4$.
More precise descriptions and statements will be given below.
The formula (\ref{anormal}) agrees with the $N=4$ result
of \cite{KG} and with the ${1\over d}$-expansion
of \cite{Weiz,Weiz2}.
\vskip 0.3cm

Following ref.\s\s\cite{KG}, we shall derive
the values of the anomalous exponents by analyzing
in perturbation theory in $\xi$ the zero modes
of differential operators characterizing
the stationary state. Although for $\xi=0$ one
observes a purely diffusive behavior of $T$
and for $\xi>0$ an inertial energy cascade,
the zero modes differ little in both cases,
the different physics arising from their cumulative
effect. As already stressed in \cite{KG},
this resembles the situation in the renormalization
group analysis in field theory or statistical
mechanics where relevant perturbations,
controlable in the single scale problem, may have
large effects on the behavior of the system.
As in the renormalization group study of critical
models, the 1$^{\rm st}$ order perturbative
corrections to the zero modes lead to the resummation
of leading infrared logarithms
in the perturbation expansion of the structure
functions in powers of $\xi$. Pursuing further
the analogy we shall introduce, as
in the perturbative renormalization group, the
notion of matrix of anomalous dimensions,
see also \cite{Weiz2}.
\vskip 0.3cm

Our results can be used to deduce the scaling properties
of the correlation functions of the dissipation field, which
we denote by $\ep(x)$, as discussed for example in
\cite{Weiz,Proc}. At finite diffusivity $\nu\not= 0$, the
dissipation field is defined by $\ep(x)= \nu
\lim_{x'\to x}\s (\nabla T)(x')\cdot(\nabla T)(x)$.
This is a sensible definition since at finite $\nu$ the
correlations of $T$ and their first derivatives are not singular
at coinciding points (the higher derivatives are).
In the limit $\nu\to 0$, we have alternative definitions:
\qq
\ep(x)\s=\s\lim_{\nu\to 0} ~\nu \lim_{x'\to x}~ (\nabla T)(x')
\cdot(\nabla T)(x)\label{diss}
\qqq
or
\qq
\ep(x)\s=\s\lim_{x'\to x}~{_1\over^2}
\({d^{\alpha\beta}(x-x')\s\d_{x^\alpha}
\d_{{x'}^\beta}}\)\s\lim_{\nu\to 0}~T(x')\s T(x)\s.
\label{dissip}
\qqq
The order of the limits in the first definition is crucial
since when $\nu\to0$ and for small $|x-x'|$,
$\ T(x)\s T(x')\s\sim\s|x-x'|^{2-\xi}$ modulo
more regular terms so that $(\nabla T)(x)\cdot(\nabla T)(x')
\s\sim\s|x-x'|^{-\xi}$ and becomes singular.
The non-commutativity of the limits $\nu\to 0$ and $x'\to x$
is at the origin of the dissipative anomaly.
The second definition of $\ep(x)$ is in the spirit
of the operator product expansion in the $\nu=0$ theory.
Using the Hopf identities (\ref{ward}) for the correlation
functions, we shall argue that both expressions
for the dissipation field $\ep(x)$ coincide for $\xi<1$.
The mean dissipation $\bar\ep\equiv\vev{\ep(x)}$
is equal to ${1\over2}C(0)$ i.e.\s\s to
the mean injection rate of energy.
The dissipation field has zero naive scaling dimension since
$T^2(x)$ and $d(x)\nabla^2_x$ have opposite naive
dimensions. However, as a consequence of the relation
(\ref{dissip}), one finds that $\ep(x)$ acquires
an anomalous scaling. In fact, the definition (\ref{dissip})
and eq.\s\s(\ref{prop1}) allow to compute any structure
functions with (non-coincident) insertions of the dissipation
field. For example, the connected 2-point function
of $\ep$ scales as
\qq
\vev{\ep(x),\ep(0)}^c\sim \({\frac{L}{|x|}}\)^{\rho_4}
\label{di}
\qqq
and it decreases with $|x|$, in agreement with
the physical picture of the dissipation
being a local process. Similarly, the $n$-point
functions of $\ep$ scale with exponents $\rho_{2n}$.
The short distance singularity in eq.\s\s(\ref{di})
is an unphysical artifact of the assumed
short distance scaling of the advecting velocity,
mollified in real systems by viscosity.
\vskip 0.3cm

The same method allows to obtain
information about the dissipative terms appearing
in the differential equations obeyed
by the structure functions and to compare our
results with the early attempts \cite{Krai94}
to calculate the anomalous exponents of the passive
scalar and with the more recent ideas \cite{Polyakov}
about the behavior of the probability distribution
functions in the turbulent systems.
\vskip 0.5cm

\section{Inertial range scaling and the zero modes}

\def\Tsl{\,\raise.15ex\hbox{/}\kern-.73em T}
The correlation functions of $T$ satisfy
the (Hopf) identities which may be deduced using
standard functional manipulations of stochastic
differential equations, see e.g. \cite{Itz}
or, for the present context, \cite{KG0}.
In the stationary state, the odd correlations vanish
and the even ones satisfy at equal times the identities
\qq
\sum_{j=1}^N \(-{\nu\De_{j}
+ {_1\over^2}D(d-1)\s\CM_N }\) \vev{T(x_1)\dots T(x_N)} \non\\
=\ \sum_{j<k} C(x_{jk}/L)
\ \vev{T(x_1)\smash{\mathop{\dots\dots}_{\hat{j}\ \
\hat{k}}}T(x_N)}
\label{ward}
\qqq
with $x_{jk}\equiv x_j-x_k\s$, $\s\De_j$ denoting the Laplacian
in the $x_j$ variable and with $\CM_N$ standing for the differential
operators given by
\qq
{_1\over^2}D(d-1)\s\CM_N =  -\s\frac{_{D(0)}}{^2}\s\s
({\s\sum_{j=1}^N \nabla_{x_j}})^2\s
+\half\sum_{j\not= k}~ d^{\alpha\beta}
(x_{jk})\s\d_{x_j^\alpha}\d_{x_k^\beta}\s\s.
\label{opMN}
\qqq
The first operator on the r.h.s. of eq.\s\s(\ref{opMN}) is zero
by translation invariance and $\CM_N$ is a sum of the 2-body
operators. For $\nu>0$, the operators appearing on the
l.h.s. of equations (\ref{ward}) are elliptic and positive.
We may use their Green functions to solve the equations inductively.
This will produce equal-time stationary correlators
decaying at infinity. Physically, they describe the stationary
state obtained by starting e.g. from a fixed localized
configuration of the scalar and waiting long enough.
\vskip 0.3cm

Notice that at $\xi=0$ the operator $\CM_N$ reduces
in the translation invariant sector to the Laplacian
in $N$ variables $x_j$:
$\CM_N|_{\xi=0}=-\De_N=-\sum_{j=1}^N \De_{j}$.
This implies that $T$ becomes
a Gaussian field at $\xi=0$ with the higher correlation
functions built in the standard way from the 2-point ones.
The stationary state coincides then with that of the
forced diffusion with the effective diffusion constant equal
to $\s\nu+\half D(d-1)$.
\vskip 0.3cm

We shall describe the inertial range correlators by taking
the limit $\nu\to 0$ at fixed positions $x_j$ and fixed large
infrared cutoff $L$. It is not important that positions
$x_j$ be disjoint as long as we do not take derivatives
of the correlators, see the remarks after
eq.\s\s(\ref{dissip}). In the limit $\nu\to0$,
the correlation functions satisfy eq.\s\s(\ref{ward}) but without
the terms $\nu\De$. These equations completely determine
the inertial range correlators up to zero modes
of operators $\CM_N$. Physically,
the zero mode contributions are fixed by the fact that
we consider the system which is the limit of the one
with positive diffusivity $\nu$. Mathematically, this means
that in order to inductively solve eqs.\s\s(\ref{ward})
we should use Green functions of the singular elliptic
operators $\CM_N$. Such Green functions are limits
of the Green functions of the non-singular operators
corresponding to the $\nu>0$ case. It has been argued in
ref.\s\s\cite{KG,Weiz,Shrai} that the zero modes of operators
$\CM_N$ effectively appear in the inertial range
correlators and give the dominant contributions
in the limit $L\to \infty$.
\vskip 0.3cm

The zero modes in question are homogeneous under dilation,
invariant by translations, rotations, parity and symmetric
under permutations of $N$ points. Since $\CM_N$ is a sum
of 2-body differential operators, zero modes of $\CM_{N-1}$
lead by symmetrization to zero modes of $\CM_N$.
More precisely, if  $f_{N-1}(x_1,\cdots,x_{N-1})$
is a zero mode of $\CM_{N-1}$, then
\qq
f_N(x_1,\dots,x_N)=\sum_{\sig\in S_N}
f_{N-1}(x_{\sig(1)},\dots,x_{\sig(N-1)}),
\label{sym}
\qqq
where the sum is over the permutations of $N$ objects,
is a symmetric zero mode of $\CM_N$. These zero modes will
never contribute to the structure functions
$\vev{\prod_j\({T(x_j)-T(y_j)}\)}$. At $\xi=0$,
the zero modes of $\CM_N=-\De_N$ are polynomials.
For any even $N>2$ there is only one "new" zero mode
of scaling dimension $N$ that cannot be expressed
as a symmetrized sum of the zero modes
of $\CM_{N-1}$. We shall denote it by $E_0$ (of course, $E_0$
is defined only up to a combination of the latter). Explicitly,
\qq
E_0(x_1,\dots,x_N)= \sum_{{{\rm pairings}~\{(l_-,l_+)\} \atop
1\leq l_-<l_+\leq N}}
\prod_{(l_-,l_+)}x_{l_-l_+}^{\s2}\ +\ [\ .\ .\ .\ ]
\label{newzero}
\qqq
where the dots $[\ .\ .\ .\ ]$ refer to quantities
which may be written as a (symmetrized) sum of functions
depending only on $N-1$ variables.
\vskip 0.3cm

The 2-point function of the scalar in the inertial
range is \cite{Krai,KG0}
\qq
\vev{T(x_1)\s T(x_2)}\ =\ const. -\s{2\s\bar\ep\over(2-\xi)
Dd(d-1)}\s |x_{12}|^{2-\xi}\s+\s\CO(L^{-2}|x_{12}|^{4-\xi})
\label{two}
\qqq
with $const. = \CO(L^{2-\xi})$. It follows that at $\xi=0$,
where $T$ becomes a Gaussian field,
\qq
\vev{T(x_1)\dots T(x_N)}\s\vert_{\xi=0}\ \cong\
c^0_N\s E_0(x_1,\dots,x_N)\ +\ [\ .\ .\ .\ ]\ ,
\label{prop}
\qqq
where $c^0_N=\({-\s\bar\ep\over Dd(d-1)}\)^{N/2}$.
The error not contained in the $[\ .\ .\ .\ ]$ terms
is bounded by $\CO(L^{-2}(max|x_{jk}|)^{N+2})$.
\vskip 0.3cm

Upon switching on positive $\xi$, the symmetric zero
modes of degree $N$ will evolve to zero modes of $\CM_{N}$
with a modified homogeneity. They may be found by
the degenerate perturbation expansion. Again, only one
of them will not come from the zero modes of $\CM_{N-1}$.
We shall call it $F_0$. Although
for $\xi$ positive, $T$ is no longer a Gaussian field,
its correlation functions may be inductively computed
from eq.\s\s(\ref{ward}). In particular it is easy to see
that the simple expressions
\qq
A_N\s\sum\limits_{1\leq j<k\leq N}
|x_{jk}|^{(2-\xi)N/2}\ , \non
\qqq
where the coefficients $\s A_N\s=\s{2(N-2)!\over(N/2)!}
\({-\s\bar\epsilon\over (2-\xi)D(d-1)
}\)^{\hspace{-0.1cm}N/2}\s\prod\limits_{l=0}^{N/2-1}
(d+(2-\xi)l)^{-1}$, \s satisfy the version
of eq.\s\s(\ref{ward}) with $\nu=0$ and $L=\infty$.
This scaling solution obviously leads to vanishing
higher structure functions and cannot give the right answer
for the inertial range correlators. The homogeneous zero
modes of the operators $\CM_N$, which enter already
at the first inductive step (the constant
in eq.\s\s(\ref{two})), modify the answer. At further
inductive steps, the previous step modifications will
induce new ones which, however, all give rise
to combinations $[\ .\ .\ .\ ]$ of functions
depending on fewer variables except, eventually,
for the terms proportional to zero modes
of $\CM_N$. If the homogeneity degree of the zero mode
is smaller than $(2-\xi){N\over2}$, the proportionality
constant will contain a compensating positive
power of $L$ and may give the contribution
dominating the large $L$ structure functions
if the zero mode is not of the $[\ .\ .\ .\ ]$ type.
Indeed, for small positive $\xi$ there is only
one non-$[\ .\ .\ .\ ]$ zero mode which we have denoted by $F_0$.
Its homogeneity degree is $(2-\xi){N\over2}-\rho_N$
with positive $\rho_N$, as will be demonstrated below.
\vskip 0.3cm

As a result, for small $\xi>0$,
\qq
\vev{T(x_1)\dots T(x_N)}\ \cong\ c_N\s L^{\rho_N}\s
F_0(x_1,\dots,x_N)\ +\ [\ .\ .\ .\ ]
\label{prop1}
\qqq
with the non-$[\ .\ .\ .\ ]$ error
bounded by $\CO(L^{-2+\CO(\xi)}
(max|x_{jk}|)^{N+2+\CO(\xi)})$. For $\xi$
not very small, the perturbations of zero modes
which at $\xi=0$ have degree higher than $N$ may
eventually enter the interval of scaling dimensions
smaller than $(2-\xi){N\over2}$ and give non-negligible
or even dominant contributions to the structure functions.
The large $L$ and $\xi\to 0$ limits of the correlation
functions of $T$ do not commute since the terms scaling
with different powers of $L$ become degenerate for
$\xi=0$, see \cite{KG}. These limits, however,
do commute for the structure functions involving
only the $F_0$ contribution scaling as $L^{\CO(\xi)}$
and the error bounded by $L^{-2+\CO(\xi)}$.
As $F_0|_{\xi=0}=E_0$, it follows by comparison
of (\ref{prop}) and (\ref{prop1}) that the
amplitude $c_N=c^0_N+\CO(\xi)$. The $\CO(\xi)$
contributions to the amplitudes $c_N$ depend on the shape
of covariance $C$ and hence are not universal.
\vskip 0.3cm

The relation (\ref{prop1}) implies the behavior (\ref{behav})
of the $N$-point structure functions $S_N(x)\equiv\vev{\s
(T(x)-T(0))^N}$. In the Gaussian limit,
\qq
S_N(x)|_{\xi=0}\s=\s a_N^0\s |x|^N
\label{sn}
\qqq
where \s$a_N^0\s=\s{N!\over(N/2)!}
\({\bar\ep\over Dd(d-1)}\)^{N/2}\s$. It follows
from the continuity of the structure functions at $\xi=0$
that the amplitude $a_N$ in eq.\s\s(\ref{behav}) is
equal to $a_N^0+\CO(\xi)$.
\vskip 0.3cm

In the perturbation expansion in powers of $\xi$,
\qq
F_0=E_0+\xi G_0 + \CO(\xi^2).
\label{o1}
\qqq
In the next sections, we shall compute the $\CO(\xi)$
contribution $G_0$ (modulo $[\ .\ .\ .\ ]$ terms).
Inserting the decomposition (\ref{o1}) into
(\ref{prop1}), we obtain an asymptotic expression
for the structure functions which, although obtained
by the first order zero-mode analysis,
contains all orders in $\xi$ resumming the
series $\sum\alpha_n\xi^n(\log L)^n$
of logarithmic infrared divergences appearing
in the expansion of the structure functions in powers
of $\xi$. This is the situation well known
from the perturbative renormalization group where
the first order approximation to the single renormalization
group step leads upon iterations to the resummation
of the leading logarithms in the perturbative expansion
of correlation functions.
\vskip 0.5cm

\section{Anomalous dimensions at $\CO(\xi)$}

Let us discuss the perturbative calculation
of the homogeneous zero modes of $\CM_N$.
At the first order in $\xi$ we have
\qq
\CM_N = - \De_N + \xi V_N + \CO(\xi^2) \non
\qqq
with $\De_N$ the Laplacian in $N$-variables and $V_N$
given by
\qq
V_N= \sum_{1\leq j\not= k \leq N}\({
\de^{\alpha\beta}\log|x_{jk}| - \inv{(d-1)}
\frac{x_{jk}^\alpha x_{jk}^\beta}{|x_{jk}|^2} }\)\d_{x^\alpha_j}
\d_{x^\beta_k}\ -\ \inv{(d-1)} \De_N\s.
\label{defV}
\qqq
Note that, since $\CM_N$ is a homogeneous operator
of dimension $\xi-2$, we have
\qq
\[\sum x_j^\alpha\d_{x^\alpha_j}\s, V_N\]= -\De_N-2V_N\s.
\label{ho}
\qqq
\vskip 0.3cm

Let $E$ be a symmetric homogeneous zero mode
of $\CM_N|_{\xi=0}$ of degree $N$.
We shall search for the zero mode of $\CM_N$ of the
form $F= E +\xi G + \CO(\xi^2)$.
The zero-mode equation gives at the order linear
in $\xi$
\qq
-\De_N G + V_N  E = 0\s.
\label{EG}
\qqq
The solutions $G$ of this equation are clearly
defined up to zero modes of $\De_N$. Note that due
to the scaling properties of $ E$ and $V_N$,
\qq
-\De_N\({\sum x_j^\alpha\d_{x^\alpha_j}-N}\)G
\s=\s -\({\sum x_j^\alpha\d_{x^\alpha_j}-N+2}\)\De_N\s G\cr
= -\({\sum x_j^\alpha\d_{x^\alpha_j}-N+2}\)V_N E
\s=\s\De_N E\s=\s0\s.
\qqq
Hence the function $E'\equiv
\({\sum x_j^\alpha\d_{x^\alpha_j}-N}\)G$
is necessarily a zero mode of $\De_N$. We shall show that
there exist solutions $G$ of eq.\s\s(\ref{EG})
such that $E'$ are homogeneous polynomials of degree $N$.
Such solutions are defined up to degree $N$ zero modes
of $\De_N$ but this ambiguity does not show up in $E'$.
We obtain this way a linear transformation $$\Ga:E\mapsto E'$$
of the space of symmetric homogeneous zero modes of $\De_N$
of degree $N$. If $(E_a)$ is a basis of this space then
the matrix $(\Ga^{a}_{\ b})$ of this transformation given by
$E'_b=\Ga^{a}_{\ b}\s E_a$ plays the role of the
{\it matrix of anomalous dimensions} at first order in $\xi$.
Indeed, if $E=v^bE_b$ is an eigenvalue $\lambda$
eigenvector of the transformation $\Ga$,
i.e. if $(v^b)$ is an eigenvector of matrix $(\Ga^{a}_{\ b})$,
then, for the corresponding
solution of eq.\s\s(\ref{EG}), we obtain
\qq
\({\sum x_j^\alpha\d_{x^\alpha_j}-N}\)G
=\lambda E
\qqq
or
\qq
\(\sum x_j^\alpha\d_{x^\alpha_j}-N-\xi\lambda\)
(E+\xi G) = \CO(\xi^2)
\qqq
which means that $E+\xi G$ is homogeneous
of order $N+\xi\lambda$ up to $\CO(\xi^2)$. Hence
the homogeneous zero modes of $\CM_N$ are perturbations
of the $\xi=0$ zero modes corresponding to eigenvectors
of the matrix of anomalous dimensions.
If the matrix $(\Ga^{a}_{\ b})$ is not totally diagonalizable
then there will be logarithmic corrections to the zero-mode
homogeneity \cite{KG0}.
\vskip 0.3cm

Reflecting the fact that all but one zero modes of $\CM_N$
are obtainable from those of $\CM_{N-1}$ by symmetrization,
the matrix of anomalous dimension is block
triangular. Namely:
\qq
(\Ga^{a}_{\ b})\s = \pmatrix{ \Ga^0_{\ 0} & 0 \cr
		    \Ga^{a'}_{\ 0} & \Ga^{a'}_{\ b'} \cr} \non
\qqq
if the matrix is written in a basis $(E_0,(E_{a'}))$ where
$E_0$ is the zero mode defined in eq.\s\s(\ref{newzero}) and
$(E_{a'})$ forms a basis of the degree $N$ zero modes arising
by symmetrization of functions depending on at most $N-1$ variables.
The matrix element $\Ga^0_{\ 0}$ is necessarily an eigenvalue
of the matrix $(\Ga^{a}_{\ b})$. If by an adequate choice
of the $[\ .\ .\ .\ ]$ terms in its definition $E_0$
becomes the corresponding eigenvector of the transformation
$\Ga$ then $F_0=E_0+\xi G_0+\CO(\xi^2)$ describes
the perturbed homogeneous zero mode of $\CM_N$ and $\Ga^0_{\ 0}$
gives the $\CO(\xi)$ correction to the scaling exponent
of $F_0$ equal to $N$ in the leading order.
The anomalous exponent of the $N$-point
structure function is therefore given by
\qq
\rho_N= -\xi(\frac{N}{2} +\Ga^0_{\ 0})\label{rhola}
\label{ga00}
\qqq
since the naive $N$-point scaling dimension
is $(2-\xi){N\over2}$.
\vskip 0.3cm

In principle it may happen (although it does not at least
for $N=2,4,6$) that $\Ga^0_{\ 0}$ is a degenerate eigenvalue
of $\Ga$ and there is no corresponding eigenvector $E_0$.
It is easy to see, however, that even in this case
there exists a zero mode $F_0=E_0+\xi G_0+\CO(\xi^2)$
of $\CM_N$ which is homogeneous of degree $N+\xi\s\Ga^0_{\ 0}
+\CO(\xi^2)$ up to $[\ .\ .\ .\ ]$ terms (the
homogeneous terms are accompanied by the ones with
powers of logarithms, the latter appearing in the $[\ .\ .\ .\ ]$
subspace). Such modifications would not effect the analysis
of the structure functions.
\vskip 0.5cm

\section{Leading order corrections to the zero modes}

Let us return to the analysis of eq.\s\s(\ref{EG}).
We shall search for the solution $G$ in the form
\qq
G\s=\s\sum_{j\not= k} \Big({H_{jk}~\log|x_{jk}| }\Big)
+ H \label{ansatz}
\qqq
with $H_{jk}$ and $H$ polynomials of degree $N$.
For such a solution,
\qq
E'\equiv\(\sum x^\alpha_j\d_{x^\alpha_j}-N\)G=\sum
\limits_{j\not=k}H_{jk}
\label{sumh}
\qqq
would necessarily be a zero mode of $\De_N$ of
degree $N$, as required. We shall see that there
indeed exist solutions of (\ref{EG}) of the form
(\ref{ansatz}) (unique up to degree $N$ zero modes
of $\De_N$) and that the polynomials $H_{jk}$
scale as $|x_{jk}|^2$ when $|x_{jk}|\to0$ assuring
that the logarithms in (\ref{ansatz}) do not lead
to divergent singularities in the correlation
functions of $T$ at coinciding points. Note, however,
that the divergences at coinciding points start to appear
in the correlation functions involving double derivatives
of $T$ or products of two first derivatives.
\vskip 0.3cm

The substitution of the Ansatz (\ref{ansatz}) into
eq.\s\s(\ref{EG}) gives a set of three equations
for $H_{jk}$ and $H$:
\qq
\De_N\s H_{jk}&=& \nabla_j\cdot\nabla_k~ E\s, \label{eqa}\\
\Big({d-2+x_{jk}\cdot\nabla_{jk}}\Big) H_{jk}
+\inv{2(d-1)} \Big({x_{jk}^\alpha x_{jk}^\beta
\d_{x^\alpha_j}\d_{x^\beta_k}}\Big) E
&=& -\half x_{jk}^2\s K_{jk}\s, \label{eqb}\\
\De_N\s H &=& \sum_{j\not= k} K_{jk} \label{eqc}
\qqq
where $K_{jk}$ are polynomials of degree $N-2$ and $\nabla_{jk}
\equiv\nabla_j-\nabla_k$ with $\nabla_j=(\d_{x^\alpha_j})$.
Eq.\s\s(\ref{eqc}) is for free since any polynomial of degree $N-2$
is in the image of $\De_N$ acting on polynomials of degree $N$.
Thus, given the solution $K_{jk}$ of eqs.\s\s(\ref{eqa},\ref{eqb}),
there always exists a degree $N$ polynomial $H$ solving
eq.\s\s(\ref{eqc}) and it is unique up to the zero modes of $\De_N$.
\vskip 0.3cm

We are thus left with solving eqs.\s\s(\ref{eqa},\ref{eqb}).
We shall first prove that there is a unique solution of
these equations and then we shall produce the solution when the
initial zero mode $E$ is the "new" zero mode $E_0$ defined by
eq.\s\s(\ref{newzero}). Notice that this is now eq.\s\s(\ref{eqa})
which implies that $\sum_{j\not= k} H_{jk}$ is a zero mode of
$\De_N$. By symmetry, we may specialize eqs.\s\s(\ref{eqa},
\ref{eqb}) to $j=1,~k=2$.
Let us work in the variables $x={x_1+x_2\over2}$, $y=x_{12}$
and $x_3,\dots,x_N$. We have:
$x_{12}\cdot\nabla_{12}=2\s y\cdot\nabla_y$.
Since $y\cdot\nabla_y$ counts the degree in $y$, we shall decompose
all terms entering in eq.\s\s(\ref{eqb}) into a sum of terms
of given degree in $y$. Namely: $H_{12}=\sum_{p=0}^N H_{12}^{(p)}$,
$K_{12}=\sum_{p=0}^{N-2} K_{12}^{(p)}$ and
\qq
-\s{1\over2(d-1)}\Big( x_{12}^\al x_{12}^\beta\s
\d_{x_1^\al}\d_{x_2^\beta}\Big)E
=\sum_{p=2}^N \tilde E^{(p)} \non
\qqq
with $H_{12}^{(p)}$, $K_{12}^{(p)}$ and $\tilde E^{(p)}$ homogeneous
polynomials in $y$ of degree $p$.
The operator $(d-2+ 2y\cdot\nabla_y)$ is
invertible on such homogeneous polynomials. Eq.\s\s(\ref{eqb})
implies then that
\qq
H_{12}^{(0)}&=& H_{12}^{(1)} = 0\s, \non\\
H_{12}^{(p)} &=& \inv{d-2 + 2p}\Big(\tilde E^{(p)}
- \half y^2 K_{12}^{(p-2)}\Big)
\quad{\rm for}\quad 2\leq p \leq N\s.\non
\qqq
The fact that $H_{12}^{(0)}= H_{12}^{(1)} = 0$ implies
that $H_{12}$ scales like $|x_{12}|^2$ when $x_1\to x_2$.
Equation (\ref{eqa}) may then be rewritten as
\qq
2\De_y H_{12}^{(p)}= (\nabla_1\cdot\nabla_2\s E)^{(p-2)}
- \De^\perp H_{12}^{(p-2)} \non
\qqq
for $p \geq 2$, where $\De^\perp=\half\De_x+\sum\limits_{j=3}^N
\De_j$. With the use of the previous relation between
$H_{12}^{(p)}$ and $y^2 K_{12}^{(p-2)}$, \s the latter equation
takes the form
\qq
\De_y\Big( y^2 K_{12}^{(p)}\Big) = f_p \non
\qqq
for some recursively known homogeneous polynomials $f_p$.
These equations may be solved for $K_{12}^{(p)}$ since
$\De_y y^2$ is an invertible operator on the space
of homogeneous polynomials of fixed degree.
\vskip 0.3cm

Let us find the deformed zero mode $F_0$ which at
$\xi=0$ reduces to $E_0$ of eq.\s\s(\ref{newzero}).
To solve eqs.\s\s(\ref{eqa},\ref{eqb}) (by symmetry,
we may again set $j=1$ and $k=2$), we first have
to compute $(\nabla_1\cdot\nabla_2) E$ and
$\Big(x_{12}^\alpha x_{12}^\beta\d_{x_1^\alpha}\d_{x_2^\beta}
\Big) E$.
\qq
(\nabla_1\cdot \nabla_2) E &=& -\s2\s (d+N-2)\s
{\sum}'\prod_{(l_-,l_+)}x_{l_-l_+}^{\s2}\ +\ [\ .\ .\ .\ ]_{12}
\label{eqea}\\
\Big(x^\alpha_{12}x^\beta_{12}\d_{x^\alpha_1}\d_{x^\beta_2}\Big)
E &=& -\s2\s x^2_{12}\s{\sum}'\prod_{(l_-,l_+)}x_{l_-l_+}^{\s2}
\non\\
&+&2\hspace{-0.3cm}\sum_{3\leq j<k\leq N}
\hspace{-0.2cm}(x_{1j}^2-x^2_{2j})(x_{1k}^2-x_{2k}^2)\s\s
{\sum}''\prod_{(l_-,l_+)}x_{l_-l_+}^{\s2}\ +\ [\ .\ .\ .\ ]_{12}
\label{eqeb}
\qqq
where ${\sum}'$ denotes the sum over pairings $\{(l_-,l_+)\}$
with $3\leq l_-<l_+\leq N$ and ${\sum}''$ a similar sum but
with, additionally, $l_\pm\not=j,k$. The symbol
$\s[\ .\ .\ .\ ]_{12}$ refers to a sum of terms which do not
depend on at least one ${x_p}$ with $p\geq 3$.
Recall from the proof of existence of solutions
of eqs.\s\s(\ref{eqa},\ref{eqb}) that $H_{12}$ has to scale
at least as $|x_{12}|^2$ as $x_1\to x_2$. It follows then that
it must be of the form
\qq
H_{12}\s=\s a~L_{12}\s +\s b~x^2_{12}\s{\sum}'
\prod_{(l_-,l_+)}x_{l_-l_+}^{\s2}\ +\ [\ .\ .\ .\ ]_{12}
\label{ansatzH}
\qqq
for some coefficients $a$ and $b$ where
\qq
L_{12}\s=\s \sum_{3\leq j<k\leq N} (x_{1j}^2-x^2_{2j})
(x_{1k}^2-x_{2k}^2)\s{\sum}''
\prod_{(l_-,l_+)}x_{l_-l_+}^{\s2}\s.
\label{defL}
\qqq
Note two properties of $L_{12}\s$:
\qq
&&(x_{12}\cdot\nabla_{12})~ L_{12}\s=\s 4\s L_{12}\s, \label{eqla}\\
&&\De L_{12}\s=\s -\s4\s(N-2)\s{\sum}'\prod_{(l_-,l_+)}x_{l_-l_+}^{\s2}
\ +\ [\ .\ .\ .\ ]_{12}\hspace{0.08cm}.
\label{eqlb}
\qqq
The first relation just means that $L_{12}$ is a homogeneous
function of $x_{12}$ of degree 2. It implies that
$$(d-2+x_{12}\cdot\nabla_{12})H_{12}= (d+2) H_{12}\ +
\ [\ .\ .\ .\ ]_{12}\hspace{0.08cm}.$$
Comparing this with the relation (\ref{eqeb}), we obtain from
eq.\s\s(\ref{eqb}) the value of the coefficient $a$:
\qq
a= -\inv{(d-1)(d+2) }\s.\non
\qqq
Next, it follows from the relation (\ref{eqlb}) that
\qq
\De H_{12} = \({\frac{4(N-2)}{(d-1)(d+2)} + 4d\s b}\)
{\sum}'\prod_{(\l_-,l_+)}x_{l_-l_+}^{\s2}\
+\ [\ .\ .\ .\ ]_{12}\s.\non
\qqq
Comparison of eqs.\s\s(\ref{eqa}) and (\ref{eqea}) gives
the value of the coefficient $b$:
\qq
b = -\inv{d}\({\frac{N-2}{(d-1)(d+2)}+ \frac{d+N-2}{2}}\).\non
\qqq
This completely determines $H_{12}$ up to terms
$[\ .\ .\ .\ ]_{12}$.
\vskip 0.3cm

Finally, in order to find the anomalous dimension $\Ga^0_{\ 0}$,
we recall that the matrix of anomalous dimensions is
found by looking at ${\sum_{j\not= k} H_{jk}}$,
cf. eq.\s\s(\ref{sumh}). $\Ga^0_{\ 0}$ is obtained by projecting
the relation (\ref{sumh}) on $E_0$ using the triangular
structure of the transformation $\Ga$.  Gathering all the terms
in eq.\s\s(\ref{ansatzH}), we obtain after a simple algebra
\qq
\sum_{j\not=k} H_{jk} &=&
-\frac{N(d+N)}{2(d+2)}
\sum_{{\rm pairings}~\{(l_-,l_+)\} \atop 1\leq l_-<l_+\leq N}
\prod_{(l_-,l_+)}x_{l_-l_+}^{\s2}\ +\ [\ .\ .\ .\ ]\non\\
&=& -\frac{N(d+N)}{2(d+2)}~ E_0\ +\ [\ .\ .\ .\ ]\hspace{0.08cm}.
\non
\qqq
Thus
\qq
\Ga^0_{\ 0}\s=\s -\frac{N(d+N)}{2(d+2)} = -\frac{N}{2}
- \frac{N(N-2)}{2(d+2)} \label{final}
\qqq
which via eq.\s\s(\ref{ga00}) leads to the claimed
value (\ref{anormal}) of the anomalous exponent $\rho_N$.
\vskip 0.5cm

\section{Dissipation field}

Let us sketch the argument for the equality of two definitions
(\ref{diss},\ref{dissip}) of the dissipation field $\ep(x)$
at $\nu=0$. First, we shall retrace the self-consistency
arguments about the short distance behavior of the
correlation functions \cite{Weiz,Proc}. These
go as follows. By separating terms, the Hopf identity
(\ref{ward}) may be rewritten in the form:
\qq
&&\(-\nu\De_1-\nu\De_2+\half D(d-1)
\CM_{(1,2)}\)\vev{T(x_1)T(x_2)T(x_3)\dots T(x_N)}\cr
&&=\ \sum\limits_{j=3}^N\(\nu\De_{j}
-\half D(d-1)(\CM_{(1,j)}+\CM_{(2,j)})\)
\vev{T(x_1)\dots T(x_N)}\hspace{0.6cm}\cr
&&-\ \half D(d-1)\hspace{-0.25cm}
\sum\limits_{3\leq j<k\leq N}\hspace{-0.25cm}\CM_{(j,k)}\s\s
\vev{T(x_1)\dots T(x_N)}\ +\ \sum_{j<k} C(x_{jk}/L)\
\vev{T(x_1)\smash{\mathop{\dots\dots}_{\hat{j}\ \
\hat{k}}}T(x_N)}\hspace{1cm}
\label{ward1}
\qqq
where the 2-point operator $\half D(d-1)\CM_{(j,k)}\equiv
d^{\alpha\beta}(x_{jk})\s\d_{x_j^\alpha}\d_{x_k^\beta}$.
In variables $x={x_1+x_2\over 2}$ and $y=x_{12}$
the l.h.s. of eq.\s\s(\ref{ward1}) becomes
\qq
\(-2\nu\De_y-d^{\alpha\beta}(y)\s\d_{y^\alpha}\d_{y^\beta}
-\half\nu\De_x+{_1\over^4}
d^{\alpha\beta}(y)\s\d_{x^\alpha}\d_{x^\beta}\)
\vev{T(x_1)\dots T(x_N)}\s.
\non
\qqq
Eq.\s\s(\ref{ward1}), with the use of the latter
decomposition and of the relation
$\d_{x^\alpha}=\d_{x_1^\alpha}+\d_{x_2^\alpha}
=-\sum\limits_{j=3}^N
\d_{x_j^\alpha}\s$, \s allows to write
\qq
\(-2\nu\De_y-d^{\alpha\beta}(y)\s\d_{y^\alpha}\d_{y^\beta}\)
\vev{T(x+\half y)T(x-\half y)T(x_3)\dots T(x_N)}\s=\s R
\label{roz}
\qqq
where $R$  is a combination of terms involving
only $x_j$-derivatives and at most first $y$-derivatives
of $\vev{T(x+\half y)T(x-\half y)T(x_3)\dots T(x_N)}$.
Let us assume that the limit when $x_{12}\to 0$
of $\vev{T(x_1)T(x_2)T(x_3)\dots T(x_N)}$ and of
$\vev{(\nabla T)(x_1)T(x_2)T(x_3)\dots T(x_N)}$
and of their derivatives over $x_j$, $j\geq 3$,
exists uniformly in small $\nu>0$ (for separated
$x,x_3,\dots x_N$). Then, as in the analysis
of the 2-point function (with anisotropic forcing),
one infers from the eq.\s\s(\ref{roz}) that
\qq
\vev{T(x+\half y)T(x-\half y)T(x_3)\dots T(x_N)}
\s=\s c_1 |y|^2(\nu+\half D(d-1)|{y}|^\xi)^{-1}\cr
+\ {\rm zero\ modes\ of\s\ }(-2\nu\De_y-d^{\alpha\beta}(y)\s
\d_{y^\alpha}\d_{y^\beta})\ +\ {\rm error}
\qqq
with the coefficients depending on $x,x_3,\dots x_N$
and the error more regular when $\nu\to0$ and $y\to0$.
The zero modes contain a polynomial of the first order
in $y$. Of the remaining zero modes the most dangerous
one comes from the angular momentum 2 sector and it behaves
like $\CO(|y|^{\alpha_2})$ for $\nu\ll\half D(d-1)|y|^\xi$,
where $\alpha_2=\half(-d+2-\xi+\sqrt{(d-2+\xi)^2
+8d})\s>\s2-\xi$, and like $\CO(|y|^2)$ for
$\half D(d-1)|y|^\xi\ll\nu$. All such terms
and their first $y$-derivatives have limits when
$y\to0$ uniformly in small $\nu$. As we see,
our assumptions about the correlators of $T$
are at least self-consistent. They are confirmed
by our $\CO(\xi)$ computation of the structure
functions. Indeed, at $\nu=0$
and for large $L$ the structure functions receive
the dominant contribution from the zero modes $F_0$
of $\CM_N$ which behave like $\xi\s
\CO(|y|^2)\log|y|$ modulo a first order polynomial in $y$
and $\CO(\xi^2)$ terms, in agreement with the
above analysis. Note that the $\xi\s\CO(|y|^2)\log|y|$
contribution to $F_0$ is not rotationally invariant
in $y$: it receives contributions from both
the $\CO(|y|^{2-\xi})$ and the $\CO(|y|^{\alpha_2})$
angular momentum 2 terms in $\vev{T(x_1)\dots T(x_N)}$.
\vskip 0.3cm

Let us use our self-consistent assumptions about
$\vev{T(x_1)\dots T(x_2)}$ in a version of eq.\s\s(\ref{roz}):
\qq
\(2\nu\s\nabla_1\cdot\nabla_2+
d^{\alpha\beta}(x_{12})\s\d_{x_1^\alpha}\d_{x_2^\beta}\)
\vev{T(x_1)T(x_2)T(x_3)\dots T(x_N)}\s=\s R'\s.
\label{roz2}
\qqq
Expression $R'$ involves only terms with at most
one derivative over $x_1$ or $x_2$. Therefore
$\lim\limits_{\nu\to0}\lim\limits_{x_{12}\to0}\s R'$
should exist and be equal to $\lim\limits_{x_{12}\to0}
\lim\limits_{\nu\to0}\s R'$. The same limits
applied to the l.h.s. of (\ref{roz2}) give, depending
on the order, the definitions
(\ref{diss}) or (\ref{dissip}) of the dissipation field
insertion $\ep(x)$, provided that $\xi>0$.
Indeed, under first ${x_{12}\to0}$ and then
$\nu\to0$ limits the $d(x_{12})\nabla_1\nabla_2$
term disappears due to vanishing of $d(x)$ at zero
while sending $\nu\to0$ before the ${x_{12}\to0}$ limit
kills the $\nu\nabla_1\cdot \nabla_2$ contribution.
Hence the equivalence of two definitions for $0<\xi<1$.
\vskip 0.3cm

By similar arguments, all three limits
$\lim\limits_{\nu\to0}$, $\lim\limits_{x_{12}\to0}$
and $\lim\limits_{\xi\to0}$ commute in the action
on $R'$. Applying them on the l.h.s. of eq.\s\s(\ref{roz2}),
we infer that at $\nu=0$
\qq
\lim\limits_{\xi\to0}\s\ep(x) = \half D(d-1)(\nabla T(x))^2
\label{ep0}
\qqq
and it describes the dissipation field of the scalar $T$
diffusing with the diffusion constant $\half D(d-1)$
and dissipating energy on long scales.
Note that the r.h.s. may be viewed as a direct
application of the second definition (\ref{dissip})
at $\xi=0$ whereas the application of the first one
(\ref{diss}) would give $\ep(x)=0$: the equivalence
of the definitions breaks down at $\xi=0$.
At $\nu=0$, the $x'\to x$ and $\xi\to0$ limits do
not commute for $(\nabla T)(x')(\nabla T)(x)$ although they
do commute for $T(x')T(x)$ or for $(\nabla T)(x')T(x)$.
This is due to the disappearance of the distinction
between the dissipative and the inertial range behavior
at $\xi=0$. A straightforward calculation employing the relation
(\ref{ep0}) shows that at non-coinciding points
\qq
\lim\limits_{L\to\infty}\s\lim\limits_{\xi\to0}~\vev{\ep(x_1),
\dots,\ep(x_n)}^c\s=
\s2^{n-1}(n-1)!\s d^{1-n}\s{\bar\ep}^n\s.
\label{epn}
\qqq
In particular, field $\ep(x)$ becomes constant
in space at $\xi=0$ and $L=\infty$, in agreement with
the physical picture of dissipation becoming a large
scale phenomenon when $\xi\to0$.
\vskip 0.3cm

The inertial range decay (\ref{di}) follows from
eq.\s\s(\ref{prop1}) with the use of the definition (\ref{dissip})
of the dissipation field and of the fact that $\vev{\ep(x)}^2
={\bar\ep}^2$ gives for large $L$ a subdominant
contribution to $\vev{\ep(x),\ep(0)}^c$. From eq.\s\s(\ref{epn})
we infer that the proportionality constant in (\ref{di}) is
equal to ${2\s\bar\ep^2\over d}+\CO(\xi)$. Similarly,
the mixed correlation functions $\s\vev{\ep(x_1)\dots\ep(x_n)
T(y_1)\dots T(y_m)}\s$ scale with the infrared cutoff as
$L^{\rho_{2n+m}}$ and with positions with exponent $(2-\xi)
{m\over 2}-\rho_{2n+m}$.
\vskip 0.5cm

\section{Equations for structure functions}

Much of the past attempts to understand the behavior of the
structure functions $S_N$ was based on the differential equations
satisfied by them \cite{Krai94}. These equations,
may be obtained from the $N$-point function equation
(\ref{ward}) in the following way.
Let $\de_j(x,y)$ denote the difference operator acting
on functions of $N$ variables $f(x_1,\dots,x_N)$ by
subtracting their values at
$x_j=x$ and $x_j=y$. $\s\de_j(x,y)$ commute for different $j$
and $\s\vev{\s(T(x)-T(y))^N}\s=\s\prod_j\de_j(x,y)\s\s
\vev{T(x_1)\dots T(x_N)}\s.$ \s Application of
$\prod_j\de_j(x,y)$
to eq.\s\s(\ref{ward}) results in the identity:
\qq
-\s d^{\alpha\beta}(x)\s\d_{x^\alpha}\d_{x^\beta}\s\s S_N(x)
\s+\s N(N-1)\s(\s C({_x\over^L})-C(0)\s)\s\s S_{N-2}(x)\s=\s J_N(x)
\label{sfe}
\qqq
with the dissipative contribution
\qq
J_N(x)&=&2\s\nu N\s\vev{\s(\De T)(x)\s(T(x)-T(0))^{N-1}}\s.
\label{jn}
\qqq
Alternatively, eq.\s\s(\ref{sfe}) may be obtained
directly from the basic stochastic differential equation
(\ref{eqT}). Despite prefactor $\nu$, the term $J_N$
does not vanish when $\nu\to0$ due to the dissipative
anomaly. Indeed, coefficient $\nu$ may be absorbed into the
insertions of the dissipation field:
\qq
J_N(x)&=&2\s\nu\s\De\s S_N(x)\s
-\s 2\s N(N-1)\s\vev{\s\ep(x)\s
(T(x)-T(0))^{N-2}}\cr\cr
&&\smash{\mathop{\longrightarrow}_{\nu\to0}}
\ \ -\s 2\s N(N-1)\s\vev{\s\ep(x)\s
(T(x)-T(0))^{N-2}}|_{\nu=0}\s.
\label{J}
\qqq
The mutual non-singularity of $\ep(x)$ and $((T(x')-T(0))^N$
at $x=x'$ and $\nu=0$, which has been assumed in the last
expression for $J_N$, may be checked by a self-consistent
analysis or directly for the perturbative solution.
\vskip 0.3cm

Our result (\ref{behav}) about the asymptotics of
the structure functions and eq.\s\s(\ref{sfe})
imply that at $\nu=0$ and for large $L$
\qq
J_N(x)\s\cong\s-\s b_N
\(L\over |x|\)^{\rho_N}|x|^{(2-\xi)(N-2)/2}
\qqq
with $b_N=D(d-1)\((2-\xi){N\over2}-\rho_N\)
\(d+(2-\xi){N-2\over2}-\rho_N\)\s a_N$\s.
\s Note that the last expression may be rewritten as
\qq
J_N(x)\s\cong\s{a_2 b_N\over a_N b_2}\s J_2\s{S_N(x)
\over S_2(x)}\s.
\label{abab}
\qqq
This relation may be
confirmed by a direct calculation of $J_N$
to order $\CO(\xi)$ from the dominant zero
mode $F_0$ contribution to the correlation
functions.
\vskip 0.3cm

In the inspiring paper \cite{Krai94}, Kraichnan
attempted to obtain anomalous exponents from eq.\s\s(\ref{sfe})
by assuming a relation similar to (\ref{abab}) but with
${a_2 b_N\over a_N b_2}={((2-\xi){N\over 2}-\rho_N)(d+
(2-\xi){N-2\over 2}-\rho_N)\over(2-\xi)d}$
replaced by ${N\over 2}$. His assumption led upon
insertion to \s\s(\ref{sfe}) to the quadratic
equation for the scaling dimension
$\zeta_N\equiv(2-\xi){N\over 2}-\rho_N$
\qq
\zeta_N(\zeta_N+d-2+\xi)=(2-\xi)d{_N\over^2}
\non
\qqq
whose solution gave the anomalous exponents $\rho_N$.
Note that the replacement of the factor
${N\over 2}$ on the r.h.s. of Kraichnan's equation
for $\zeta_N$ by ${a_2 b_N\over a_N b_2}$ leads instead
to a tautological identity.
\vskip 0.3cm

Kraichnan's values of $\rho_N$, unlike the ones
obtained in the present work, do
not vanish at $\xi=0$. The latter might seem strange
in view of the fact that $T$ becomes a Gaussian
field at $\xi=0$ and the Ansatz of \cite{Krai94} was fit
with the Gaussian calculation. The latter was based,
however, on the definition (\ref{jn}) applied directly
in the Gaussian case whereas at $\xi=0$ one should use
for $J_N$ the expression (\ref{J}) with $\ep(x)$
given by the equation (\ref{ep0}). The latter calculation
agrees, of course, with eq.\s\s(\ref{abab}):
\qq
J_N(x)|_{\xi=0}\s=\s -b_N^0\s |x|^{N-2}
\qqq
where $b_N^0=D(d-1)N(d+N-2)\s a_N^0$, see eq.\s\s(\ref{sn}).
In this limit, the differential equation (\ref{sfe}) reduces
to $-D(d-1)\s\De S_N|_{\xi=0}=J_N|_{\xi=0}$.
\vskip 0.3cm

Above, we have studied the inertial range behavior
of the structure functions.
More general objects to study are the joint
probability distribution functions (p.d.f.'s)
$P_N(T_1,\dots,T_N;$ $x_1,\dots,x_N)$ of the scalar whose
moments give the equal time correlation functions.
In particular, the structure functions
$\vev{\prod_j\({T(x_j)-T(y_j)}\)}$ are special moments
of the p.d.f.'s
\qq
Q_N(T_1,\dots,T_N;x_1,\dots,x_N)\s=\s\int
P_N(T_1+\tau,\dots,T_N+\tau\s;\s x_1,\dots,x_N)\ d\tau
\qqq
which are translational invariant in the $T$-variables.
The generating function of $S_N$'s,
\qq
Z(\lambda;x)\s\equiv\s\vev{{\rm e}^{\s i\lambda(T(x)-T(0))}}
\s=\s\int {\rm e}^{\s i\lambda T}\s Q(T;x)\s dT
\qqq
is a Fourier transform of the p.d.f.
$Q(T_1-T_2;x_{12})\equiv Q_2(T_1,T_2;x_1,x_2)$,
\vskip 0.3cm

In a recent paper \cite{Polyakov} on the Burgers equation,
Polyakov has argued that the structure function p.d.f.'s
exhibit a universal inertial range behavior for
(translating his statements to the passive scalar case)
$T_i\ll T_{\rm rms}$ where
$T_{\rm rms}\equiv\sqrt{\vev{T(0)^2}}=\CO(L^{1-\xi/2})$,
see eq.\s\s(\ref{two}). Polyakov's analysis was
based on postulating an operator product expansion
which allows to close the resummed version of
eq.\s\s(\ref{sfe})
\qq
-\s d^{\alpha\beta}(x)\s\d_{x^\alpha}
\d_{x^\beta}\s Z(\lambda;x)
\s+\s\la^2 \s(\s C(0)-C({_x\over^L})\s)\s\s
Z(\lambda;x)\s=\s J(\la;x)
\label{sfe1}
\qqq
by expressing its r.h.s. $J(\la;x)\equiv\sum_N{(i\la)^N
\over N!}J_N(x)=2\la^2\vev{\s\ep(x)\s
{\rm e}^{\s i\la(T(x)-T(0))}}\s$ again in terms of $Z(\la;x)$.
The resulting equation for $Z$ may be reduced
to an ordinary differential equation
by imposing the scaling relation
\qq
\(\s2\s x\cdot\nabla-(2-\xi)\s\la\d_\la\)\s Z(\la;x)\s=\s0\s\s,
\label{scal}
\qqq
i.e. by postulating that $Z(\la;x)$ depends on $\la^2|x|^{2-\xi}$.
Note that a strictly scaling solution for $Z(\la;x)$ implies
either the Kolmogorov scaling or divergence of the structure
functions. For example, the resummation of the expressions
(\ref{sn}) for the $\xi=0,\ \nu=0,\ L=\infty$ structure
functions gives the Gaussian generating function
\qq
Z(\la;x)|_{\xi=0}\s=\s \exp\[-\s{\bar\ep\over Dd(d-1)}\s\la^2
|x|^2\]\s,
\qqq
which scales, in accordance with the normal scaling
of the $\xi=0$ structure functions.
It corresponds to the Gaussian p.d.f.
\qq
Q(T;x)|_{\xi=0}\s=\s\({Dd(d-1)\over 4\pi\bar\ep}\)^{1/2}
\frac{1}{|x|}\ \exp\[-\s{Dd(d-1)
\over 4\s\bar\ep}\s{T^2\over |x|^2}\]\s.
\qqq
A straightforward check shows that
at $\xi=0$ the dissipative term $J$ may be simply
expressed in terms of $Z$ itself:
$J=2\s\bar\ep\s\la^2 (1+{1\over d}\s\la\d_\la)\s Z$.

\vskip 0.3cm

Our small $\xi$ analysis of the passive scalar
does not allow us to confirm or to infirm Polyakov's picture
since the individual structure functions that we study
probe the small $\lambda$ behavior of $Z$ i.e. the large $T$
behavior of $Q(T)$. What follows from it is that
the large $T$ tails of the p.d.f.'s violate scaling for $\xi>0$.
In particular, the solution (\ref{behav},\ref{anormal})
for $S_N$'s leads to the relation
\qq
\(2\s x\cdot\nabla-(2-{\xi\s d\over d+2})\s\la\d_\la
+{\xi\over d+2}\s(\la\d_\la)^2\) Z(\la;x)\s=\s\CO(\xi^2)
\qqq
and to the asymptotic scale $L$ dependence
\qq
\(L\d_L-{\xi\over2(d+2)}\s\la\d_\la(\la\d_\la-2)\)
Z(\la;x)\s=\s\CO(\xi^2)
\qqq
consistent, of course, with the exact overall scaling
property of the problem:
\qq
\(2\s x\cdot\nabla+2\s L\d_L-(2-\xi)\s\la\d_\la\)
Z(\la;x)\s=\s0\s.\non
\qqq
\vskip 0.3cm

It would be interesting to recover the tail of
the structure-function p.d.f. $Q$ in the
order $\CO(\xi)$. This would require the knowledge
of the $\CO(\xi)$ contributions to the non-universal
amplitudes $a_N$ in the relation (\ref{behav})
which we have not computed. It is possible that they
may be found by a perturbative analysis of
instanton contributions to the functional integral
for the dynamical scalar correlators \cite{MigFalk}.
We expect that the refined perturbative analysis in $\xi$
will allow a better control of both the structure
functions and the corresponding p.d.f.'s.

\end{document}